\documentclass[aps, prl, twocolumn, showpacs, amsmath,amssymb, superscriptaddress]{revtex4}% Physical Review L

\usepackage{graphicx}  % needed for figures
\usepackage{dcolumn}   % needed for some tables
\usepackage{bm}        % for math
\usepackage{amssymb}   % for math

\begin{document}
\newcommand{\mb}[1]{\mathbf{#1}}
\newcommand{\phipp}{\big|\phi_{\mb{p}}^{(+)}\big>}
\newcommand{\phipav}{\big|\phi_{\mb{p}}^{\p{av}}\big>}
\newcommand{\pp}[1]{\big|\psi_{p}(#1)\big>}
\newcommand{\drdy}[1]{\sqrt{-R'(#1)}}
\newcommand{\rbf}{$^{85}$Rb}
\newcommand{\rbs}{$^{87}$Rb}
\newcommand{\kf}{$^{40}$K}
\newcommand{\na}{${^{23}}$Na}
\newcommand{\muK}{\:\mu\textrm{K}}
\newcommand{\p}[1]{\textrm{#1}}
\newcommand\T{\rule{0pt}{2.6ex}}
\newcommand\B{\rule[-1.2ex]{0pt}{0pt}}
\newcommand{\reffig}[1]{\mbox{Fig.~\ref{#1}}}
\newcommand{\refeq}[1]{\mbox{Eq.~(\ref{#1})}}
\hyphenation{Fesh-bach}

\preprint{}

\title{Prediction of Feshbach resonances from three input parameters}

  \author{Thomas M. Hanna} 
  \affiliation{Atomic Physics Division,
  National Institute of Standards and Technology, 
  }

  \author{Eite Tiesinga} 
  \affiliation{Joint Quantum Institute,
  University of Maryland and NIST, 
  100 Bureau Drive Stop 8423,
  Gaithersburg MD 20899-8423, USA.}

  \author{Paul S. Julienne}
    \affiliation{Joint Quantum Institute,
  University of Maryland and NIST,
  100 Bureau Drive Stop 8423,
  Gaithersburg MD 20899-8423, USA.}

\begin{abstract}
We have developed a model of Feshbach resonances in gases of ultracold alkali metal atoms using the ideas of multichannel quantum defect theory. 
Our model requires just three parameters describing the interactions - the singlet and triplet scattering lengths, and the long range van der Waals coefficient - in addition to known atomic properties. 
Without using any further details of the interactions, our approach can accurately predict the locations of resonances.  
It can also be used to find the singlet and triplet scattering lengths from measured resonance data.
We apply our technique to 
$^{6}$Li--$^{40}$K and $^{40}$K--$^{87}$Rb scattering, obtaining good agreement with experimental results, and with the more computationally intensive coupled channels technique.
\end{abstract}

\date{\today}

\maketitle

The use of magnetic fields to control and resonantly enhance interactions in ultracold atomic gases allows the creation of highly rovibrationally excited molecules~\cite{review}.
Alongside experimental efforts, substantial work has been invested in the theory of Feshbach resonances and molecules.
The coupled channels technique~\cite{stoof88, tiesinga93, mies96} has had considerable success, but can be computationally intense.
Along with the rapidly growing number of experimentally relevant collision partners~\cite{inouye04, ferlaino06, stan04, deh08, pilch08, wille08}, this creates a need for accurate, computationally simple models of Fesh\-bach resonances.
Previous models motivated by this need have been based on, for example, an expansion in terms of the bound states of the singlet and triplet Born-Oppenheimer (BO) potentials, as with the very successful Asymptotic Bound State Model (ABM) ~\cite{wille08},
or a fixed phase at an inner radius which is matched to the long range evolution of the wavefunction~\cite{vankempen02}.

In this paper, we show how multichannel quantum defect theory (MQDT) can be used to construct a model of Feshbach resonant collisions in alkali gases that is both physically motivated and computationally simple. 
MQDT~\cite{greene82, seaton83, mies84} has been used to study collisions in a range of systems (see, e.g., \cite{idziaszek09} and references therein), and is built around the idea of a separation of energy scales. 
This applies to the scattering of ultracold alkali metal atoms because the hyperfine splitting and collision energies are much smaller than the potential depths. 
We will show that this enables a complete model of Feshbach resonances to be constructed using only the scattering lengths of the singlet and triplet BO potentials, and the common van der Waals coefficient of their long range tail. 
This represents a substantial simplification over having to make use of whole potentials, yet leaves us with an approach powerful enough to make useful predictions.
In addition, our method is computationally simple enough for these three parameters to be optimized, or even found, from measured resonance data.

A colliding pair of atoms can be described in terms of channels $|\alpha\rangle$, defined by the state of each atom, as well as the partial wave of the collision, $\ell$, and its projection onto a quantization axis, $m_\ell$. 
The channel energy $E_\alpha$ is given by the energy of the two atoms at asymptotically large separation.
Typically, multiple channels are involved, each with its own associated potential and couplings to other channels.
We use van der Waals potentials of the form $V^{\alpha}_\p{vdW}(r) = E_\alpha + \hbar^2 \ell(\ell+1)/(2\mu r^2) - C_6/r^6$, where the second term represents the centrifugal barrier, $C_6$ is the van der Waals coefficient, $\mu$ is the reduced mass, and $r$ is the interparticle separation. 
This approximation is physically motivated by the common long range form of the BO potentials, which converge to the form $-C_6/r^6$ beyond a certain radius $r^*$. 
At $r^*$, the energy scale of the potential is much greater than the channel and collision energies.
The physics occurring in the short range region $r < r^*$ can then be accounted for by imposing appropriate boundary conditions at $r^*$.
In fact, the separation of energy scales, and accompanying length scale separation, allow us to make the approximation $r^* \rightarrow 0$ and use $V^{\alpha}_\p{vdW}(r)$ at all $r$.

We calculate a pair of linearly independent reference functions, $f$ and $g$, from each potential by solving the single channel radial Schr\"odinger equation for a chosen total energy, $E$. A channel $|\alpha\rangle$ is described as open when $E > E_\alpha$, and closed when $E < E_\alpha$.
An observation of quantum defect theory is that $f$ and $g$ can be approximated to be independent of energy and angular momentum at short range.
They can be calculated numerically~\cite{julienne89, mies00b, raoult04} or, for some potentials such as $V^{\alpha}_\p{vdW}(r)$, found analytically~\cite{gao96, gao98, gao04b, gao05}.
We use the analytic approach.

We next note that, at short range, the splitting between the BO potentials is also far greater than the hyperfine, Zeeman, and collision energies.
We can therefore neglect these three contributions in this region.
Consequently, the radial multichannel scattering wavefunction can be written in the form $\vec{\psi}(r) = \vec{f}(r) - \mb{K}^{(s)} \vec{g}(r)$, where $\vec{f}(r)$ and $\vec{g}(r)$ are vectors containing the reference functions for each channel, and bold font indicates a matrix.
The short range $K$ matrix, $\mb{K}^{(s)}$, is to a good approximation independent of $r$ and $E$.

Neglecting the hyperfine and Zeeman interactions at short range also makes $\mb{K}^{(s)}$ diagonal in a basis in which the BO potentials are diagonal.
We refer to the $K$ matrix expressed in this basis as $\mb{K}^{(BO)}$.
For alkali atoms in the $^2$S electronic ground state there are several ways to construct such a basis.
%is one singlet and one triplet BO potential. 
A convenient choice is to first couple together the electron spins of the two atoms to form $\vec{S} = \vec{s}_1 + \vec{s}_2$. 
Alkalis have one singlet and one triplet BO state, which are described by $S = 0$ and 1, respectively. $\vec{S}$ is then coupled to $\vec{\ell}$ to give $\vec{J} = \vec{S} + \vec{\ell}$, which is finally coupled to the sum of nuclear spins, $\vec{I} = \vec{i}_1 + \vec{i}_2$, to give the total angular momentum, $\vec{T}$. 
This gives the kets $|(S \ell)J,I;T M_T\rangle$, which we refer to as the molecular basis. The projection $M_T$ is taken along the magnetic field axis.

The entries of $\mb{K}^{(BO)}$ depend only on whether a channel is of singlet or triplet symmetry, and are given in terms of the scattering lengths $a_\p{s,t}$ of the corresponding potentials by~\cite{gao05}:
\begin{align}
  a_{\p{s,t}} / \bar{a} = \sqrt{2} 
  \frac{K^{(BO)}_{\p{s,t}} + \tan(\pi/8)}{K^{(BO)}_{\p{s,t}} - \tan(\pi/8)} \, .
\end{align}
Here, $K^{(BO)}_{\p{s,t}}$ is the $K$ matrix element of the singlet (s) and triplet (t) channels, $\bar{a} = 2^{-3/2}[\Gamma(3/4)/\Gamma(5/4)](2\mu C_6 / \hbar^2)^{1/4}$ is the mean scattering length, and $\Gamma(z)$ is the gamma function. 

The channel basis $|\alpha\rangle$, by contrast to the molecular basis, is constructed in terms of uncoupled atomic states, plus the partial wave of their collision.
At nonzero magnetic field the projection of the total atomic angular momentum $\vec{f}_{1,2} = \vec{i}_{1,2} + \vec{s}_{1,2}$ onto the magnetic field quantisation axis, $m_{1,2}$, is a good quantum number, but $f_i$ itself is not. The states are then described by the kets
$ |\alpha_1 m_1, \alpha_2 m_2, \ell m_\ell\rangle$, which we refer to as the Zeeman basis. 
Here, $\alpha_{1,2} = a,b,c\ldots$ label the energy ordered atomic Zeeman states~\cite{review}.

We obtain $\mb{K}^{(s)}$ using a frame transformation~\cite{fano70, rau71} from the molecular basis to the Zeeman basis. This magnetic-field dependent, unitary transformation can be calculated in terms of Clebsch-Gordan coefficients and the Wigner $6j$ and $9j$ symbols using angular momentum algebra~\cite{rose}. For the case of identical atoms, it is simple to incorporate the relevant Bose/Fermi symmetrisation. Writing the frame transformation as a matrix $\mb{U}$, we have $ \mb{K}^{(s)} = \mb{ U K}^{(BO)} \mb{U}^\dagger $.

We next calculate the long range $K$ matrix, $\mb{K}(E)$, from which the observable scattering properties can be extracted. Unlike $\mb{K}^{(s)}$, the long range $K$ matrix is energy dependent because of the van der Waals tail of the potentials.
For a multichannel problem with both open and closed channels, $\mb{K}(E)$ is given by~\cite{gao05}
\begin{align}
  \mb{K}(E) &= - [\mb{Z}_{fg}(E) - \mb{Z}_{gg}(E) \mb{K}_\p{eff}^{(s)} ] \nonumber \\
  & \times [\mb{Z}_{ff}(E) - \mb{Z}_{gf}(E) \mb{K}_\p{eff}^{(s)} ]^{-1} \, ,
  \label{eq:Keff}
\end{align}
where
\begin{align}
 \mb{K}_\p{eff}^{(s)} = \mb{K}_{oo}^{(s)}+ \mb{K}_{oc}^{(s)} [\boldsymbol{\chi}(E) - \mb{K}_{cc}^{(s)}]^{-1} \mb{K}_{co}^{(s)}
\, .
\label{eq:Keffz}
\end{align}
Here, the subscripts `o' and `c' refer to the open and closed channel blocks of the $K$ matrix. 
The energy dependent, diagonal $Z$ matrices are given by the long range behavior of the reference functions $f$ and $g$, and represent the propagation of a wavefunction from small to large distances. 
The diagonal matrix of the bound state phase in the closed channels, $\boldsymbol{\chi}(E)$, is analogously defined~\cite{gao05} and, as shown by the structure of \refeq{eq:Keffz}, gives rise to the resonant enhancement of the collisions. It also 
allows the calculation of bound state energies from the determinental equation
\begin{align}
 \p{det}( \boldsymbol{\chi}(E) - \mb{K}_{cc}^{(s)} ) = 0 \, .
 \label{eq:bnd_det}
\end{align}
The $Z$ and $\chi$ matrix entries are evaluated for each channel $|\alpha\rangle$ at the magnetic field dependent energy $E - E_{\alpha}(B)$.
Finally, the $S$ matrix, which we use to extract the observable scattering properties, may be calculated from $\mb{S}(E) = [\mb{I} + i\mb{K}(E)][\mb{I} - i\mb{K}(E)]^{-1}$, where $\mb{I}$ is the identity matrix. 
%The magnetic field dependence of the scattering and bound state properties are encapsulated in the channel energies and frame transformation.
By approximating the short range interactions to be energy independent, and using the form $V^{\alpha}_\p{vdW}(r)$ for the potentials, we have therefore reduced the scattering problem to one involving just three parameters: $a_\p{s}$, $a_\p{t}$, and $C_6$.

Our code is sufficiently fast that it is possible to search over the whole $(a_\p{s}, a_\p{t})$ plane. 
Given one field at which a resonance occurs, and assuming that $C_6$ is known, lines through the $(a_\p{s}, a_\p{t})$ plane are typically found which result in a resonance at this field. 
Further resonances reduce the possible range for each scattering length.
For more accurate determinations, least squares minimization using all available resonance locations then gives optimal values of $a_\p{s}$ and $a_\p{t}$.
We note that the values thus found can be different to those produced by a full coupled channels calculation using realistic potentials, and should be interpreted as fit parameters that will in general be close to the real scattering lengths.
Nonetheless, the scattering lengths found can be used for the prediction of unobserved resonances. This could make our technique useful for investigating new collision partners.

\begin{table}[t]
\caption{Resonance locations $B_0$ and widths $\Delta B$ obtained from an MQDT
  calculation for $^{6}$Li--$^{40}$K, after varying $a_\p{s}$ and $a_\p{t}$ to optimally fit measured resonances. Experimental (Exp) and coupled channels (CC) data are from Ref.~\cite{wille08}. Channels are labelled by the Zeeman
  state of each atom, with $\alpha_1$  referring to $^{6}$Li and $\alpha_2$ referring to $^{40}$K, and the total angular momentum projection, $M_\p{T}$. `$p$' indicates a $p$ wave resonance, for which $\Delta B$ is not defined.}
\centering
\begin{tabular}{c c |c c | c | c  c } 
\hline\hline  
 &\T & \multicolumn{2}{c|}{MQDT} & Exp & \multicolumn{2}{c}{CC} \\
$\alpha_1 \alpha_2 $ & $M_\p{T}$ & $B_0$ [mT] & $\Delta B$ [mT]  
 & $B_0$ [mT]  & $B_0$ [mT]  &  $\Delta B$ [mT] \\ [1ex]
\hline 
$ba$\T & -5 & 21.36 & 0.028 & 21.56 & 21.56 & 0.025 \\
$aa$ & -4 & 15.93 & 0.022  & 15.76 & 15.82 & 0.015\\ 
$aa$ & -4 & 17.01 & 0.007 & 16.82 & 16.82 & 0.01 \\
$aa$ & -4 & 25.80 & $p$    & 24.9   & 24.95 & $p$ \\
$ab$ & -3 & 0.80   & $p$    & 1.61  & 1.05 & $p$ \\
$ab$ & -3 & 15.31 & 0.042 & 14.92 & 15.02 & 0.028 \\
$ab$ & -3 & 15.96 & 0.016 & 15.95 & 15.96 & 0.045\\
$ab$ & -3 & 16.86 & 0.005 & 16.59 & 16.59 & 0.0001 \\
$ab$ & -3 & 26.08 & $p$    & 26.3   & 26.20 & $p$  \\
$ac$ & -2 & 14.77 & 0.044 & 14.17 & 14.30 & 0.036 \\
$ac$ & -2 & 15.81 & 0.056 & 15.49 & 15.51 & 0.081 \\
$ac$ & -2 & 16.72 & 0.005 & 16.27 & 16.29 & 0.060 \\
$ac$ & -2 & 26.10 & $p$    & 27.1   & 27.15 & $p$  \\
\hline 
\end{tabular}
\label{table:lik} 
\end{table}
As an example application of our code we consider cold collisions of
$^{6}$Li and $^{40}$K atoms. 
The analysis of Ref.~\cite{wille08} found $a_\p{s} = 52.1$\,$a_0$, $a_\p{t} = 63.5$\,$a_0$, and $C_6 = 2322\, E_\p{h}a_0^6$. 
Here, $a_0 = 0.0529177$\,nm is the Bohr radius, and $E_\p{h} = 4.35974 \times 10^{-18}$\,J. 
With these input parameters, we find $s$-wave resonances in the $aa$ channel at 17.08\,mT and 18.13\,mT. 
The locations found differ significantly from the measured values of 15.76\,mT and 16.82\,mT~\cite{wille08}. 
However, allowing the singlet and triplet scattering lengths to vary, we find optimal agreement for $a_{\p{s}} = 53.17 \, a_0$ and $a_{\p{t}} = 64.41 \, a_0$. 
These values were obtained using a least squares minimization, comparing our results to the experimentally observed locations of the thirteen resonances listed in Table~\ref{table:lik}. 
Our results, obtained from only the given $C_6$ and by varying the scattering lengths, agree well with the more computationally intensive coupled channels approach, and with the experimental results.
The abovementioned resonances, for example, are found at 15.93\,mT and 17.01\,mT, a disagreement of approximately 1\,\%. 

\begin{figure}[tbp]
	\centering
\includegraphics[width=0.99\columnwidth, clip]{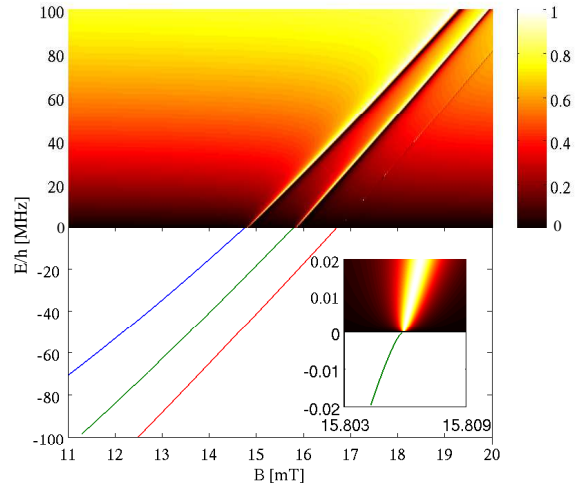}
	\caption{(color online). Bound state energies (lower panel) and scattering
	phase shift $\sin^2\delta_0$ (upper panel) for the $ac$ channel
	of $^{6}$Li--$^{40}$K, as a function of magnetic field, calculated from our three parameter MQDT model. 
	The inset shows a close-up of the 15.81\,mT resonance, illustrating the narrow region within which the Feshbach molecular state has an appreciable entrance channel component~\cite{review}.}
	\label{fig:lik}
\end{figure}

The $s$-wave resonances in the $ac$ channel listed in Table~\ref{table:lik}
are illustrated in \reffig{fig:lik}. The two panels show the effect of a Feshbach state below and above the threshold at which it causes a resonance. 
The lower panel shows the bound state energies as a function of magnetic field.
The upper panel shows $\sin^2 \delta_0 (E)$, where $\delta_0 (E)$ is the $s$-wave scattering phase shift in the $ac$ channel. 
In this example, there is one open channel and eleven closed channels. 
Consequently, the $s$-wave phase shift is linked to the $S$ matrix by $S(E) = e^{2 i \delta_0 (E)}$. 
Resonances can be recognised by the sudden change in the value of $\sin^2 \delta_0 (E)$. 

\begin{figure}[tbp]
        \centering
\includegraphics[width=0.99\columnwidth, clip]{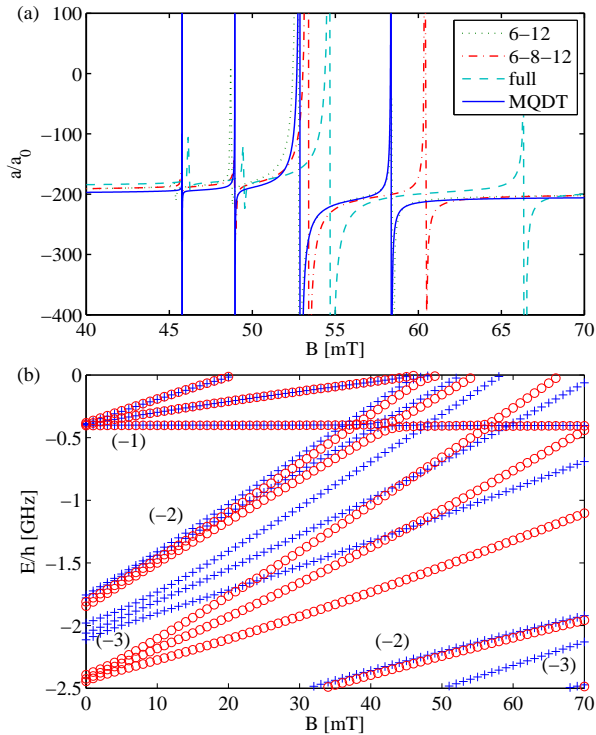}
\caption{(color online). Comparison of $^{40}$K--$^{87}$Rb $aa$ resonance locations predicted by different approaches. The upper panel shows scattering length vs magnetic field. The MQDT results are given by the approach presented in this paper. The other results are given by coupled channels calculations with potentials of the form $-C_6/r^6 + C_{12}/r^{12}$ (6--12), $-C_6/r^6 -C_8/r^8 + C_{12}/r^{12}$ (6--8--12), and the potentials of Ref.~\cite{pashov07} (full). The full potentials were also used to calculate the bound state spectrum (`o') which is compared to an MQDT calculation (`$+$') in the lower panel. Bound states are numbered according to the vibrational quantum number of the singlet and triplet states which give rise to them. Note that the two plots have different ranges of magnetic field.}
        \label{fig:krbcomp}
\end{figure}

As a second example we consider $^{40}$K$-^{87}$Rb scattering.
In \reffig{fig:krbcomp}a we compare an MQDT calculation of $s$-wave $aa$ channel resonances to three coupled channels calculations using two different pairs of singlet and triplet power law potentials, and the potentials of Pashov \textit{et al.}~\cite{pashov07}, which to the best of our knowledge are the most accurate available. 
The power-law potentials are constructed to have the same scattering lengths, $C_6$, and number of bound states as the potentials of Ref.~\cite{pashov07}. 
These values for the scattering lengths and $C_6$ are also used in the MQDT calculation, without varying their values as in the example above, to allow direct comparison.
The resulting resonance locations illustrate the two main approximations of our approach. 
Firstly, we approximate the short range $K$ matrix to be energy independent.
The validity of this is shown by the close agreement of the MQDT result to that of the coupled channels calculation with $-C_6/r^6 + C_{12}/r^{12}$ potentials.
These quickly converge to pure van der Waals potentials as $r$ increases,
only having significant $C_{12}$ contributions at distances where the hyperfine energies are small compared to $C_6/r^6$.

Our second important approximation is the use of a pure van der Waals potential. 
Other dispersion contributions to the potential, such as the attractive $-C_8/r^8$ term, can be important, as shown in \reffig{fig:krbcomp}a. 
The $-C_8/r^8$ term, taken from Ref.~\cite{pashov07}, has a sufficiently long range to give deviations from the pure van der Waals result. These results are closer to those given by the full potential.
MQDT using numerical reference functions based on a more detailed potential~\cite{vankempen02} could be more accurate, but would be more computationally involved.

Another way of illustrating the differences between the calculations is by studying the near-threshold bound states, which we show for the $aa$ channel in \reffig{fig:krbcomp}b.
The difference between the MQDT result and that of the coupled channels calculation using the potential of Ref.~\cite{pashov07} is primarily due to the zero field bound state locations being different, which is a consequence of the different singlet and triplet potentials. 
The states around $-0.4$\,GHz, arising from the most weakly bound ($v = -1$) level, agree to approximately 1\%, whereas the more deeply bound $v=-2$ and $-3$ states deviate more significantly.
The MQDT bound state energies become less accurate as binding energy increases, as non van der Waals contributions become more important.
This implies that resonances due to more deeply bound levels will be less accurately reproduced by the MQDT model. 
The fit procedure discussed above partially compensates for this. 
We note that the ABM method~\cite{wille08} has been shown to accurately predict the locations of $^{40}$K$-^{87}$Rb resonances~\cite{tiecke_private}.

The MQDT model can give a qualitative indication of resonance widths, as shown in Table~\ref{table:lik}. 
In coupled channels calculations, resonance widths are determined primarily by the difference between the singlet and triplet potentials. This is not included in the MQDT model, where the only difference between the two is in their scattering lengths.
Lastly, we note that our MQDT approach neglects the weak magnetic dipole-dipole and second-order spin-orbit interactions, which could be included in a more general MQDT treatment.

In conclusion, we have developed a model of Feshbach resonances in ultracold alkali gases utilising the separation of energy scales suggested by MQDT. 
A frame transformation allows us to link short range, where we can neglect the energy dependence of the interactions, to long range, where we can use an approximate potential with analytic reference functions.
We are therefore able to predict and study resonances on the basis of three input parameters - $a_\p{s}$, $a_\p{t}$ and $C_6$ - in addition to known atomic properties.
Our approach is significantly less computationally challenging than the coupled channels approach and, by using the scattering lengths as fit parameters, could enable useful predictions to be made while investigating new collision partners.  

We gratefully acknowledge Bo Gao for supplying us with his codes for calculating the $\mb{Z}(E)$ and $\boldsymbol{\chi}(E)$ matrices. P.S.J acknowledges partial support by the U.S. Office of Naval Research.

\bibliography{tomsrefs}

\begin{thebibliography}{27}
\expandafter\ifx\csname natexlab\endcsname\relax\def\natexlab#1{#1}\fi
\expandafter\ifx\csname bibnamefont\endcsname\relax
  \def\bibnamefont#1{#1}\fi
\expandafter\ifx\csname bibfnamefont\endcsname\relax
  \def\bibfnamefont#1{#1}\fi
\expandafter\ifx\csname citenamefont\endcsname\relax
  \def\citenamefont#1{#1}\fi
\expandafter\ifx\csname url\endcsname\relax
  \def\url#1{\texttt{#1}}\fi
\expandafter\ifx\csname urlprefix\endcsname\relax\def\urlprefix{URL }\fi
\providecommand{\bibinfo}[2]{#2}
\providecommand{\eprint}[2][]{\url{#2}}

\bibitem[{\citenamefont{{K}\"ohler et~al.}(2006)\citenamefont{{K}\"ohler,
  {G}\'oral, and {J}ulienne}}]{review}
\bibinfo{author}{\bibfnamefont{T.}~\bibnamefont{{K}\"ohler}},
  \bibinfo{author}{\bibfnamefont{K.}~\bibnamefont{{G}\'oral}},
  \bibnamefont{and} \bibinfo{author}{\bibfnamefont{P.~S.}
  \bibnamefont{{J}ulienne}}, \bibinfo{journal}{Rev. {M}od. {P}hys.}
  \textbf{\bibinfo{volume}{78}}, \bibinfo{pages}{1311} (\bibinfo{year}{2006}).

\bibitem[{\citenamefont{Stoof et~al.}(1988)\citenamefont{Stoof, {K}oelman, and
  {V}erhaar}}]{stoof88}
\bibinfo{author}{\bibfnamefont{H.~T.~C.} \bibnamefont{Stoof}},
  \bibinfo{author}{\bibfnamefont{J.~M. V.~A.} \bibnamefont{{K}oelman}},
  \bibnamefont{and} \bibinfo{author}{\bibfnamefont{B.~J.}
  \bibnamefont{{V}erhaar}}, \bibinfo{journal}{Phys. {R}ev. {B}}
  \textbf{\bibinfo{volume}{38}}, \bibinfo{pages}{4688} (\bibinfo{year}{1988}).

\bibitem[{\citenamefont{Tiesinga et~al.}(1993)\citenamefont{Tiesinga,
  {V}erhaar, and {S}toof}}]{tiesinga93}
\bibinfo{author}{\bibfnamefont{E.}~\bibnamefont{Tiesinga}},
  \bibinfo{author}{\bibfnamefont{B.~J.} \bibnamefont{{V}erhaar}},
  \bibnamefont{and} \bibinfo{author}{\bibfnamefont{H.~T.~C.}
  \bibnamefont{{S}toof}}, \bibinfo{journal}{Phys. {R}ev. {A}}
  \textbf{\bibinfo{volume}{47}}, \bibinfo{pages}{4114} (\bibinfo{year}{1993}).

\bibitem[{\citenamefont{Mies et~al.}(1996)\citenamefont{Mies, {W}illiams,
  {J}ulienne, and {K}rauss}}]{mies96}
\bibinfo{author}{\bibfnamefont{F.~H.} \bibnamefont{Mies}},
  \bibinfo{author}{\bibfnamefont{C.~J.} \bibnamefont{{W}illiams}},
  \bibinfo{author}{\bibfnamefont{P.~S.} \bibnamefont{{J}ulienne}},
  \bibnamefont{and} \bibinfo{author}{\bibfnamefont{M.}~\bibnamefont{{K}rauss}},
  \bibinfo{journal}{J. {R}es. {N}atl. {I}nst. {S}tand. {T}echnol.}
  \textbf{\bibinfo{volume}{101}}, \bibinfo{pages}{521} (\bibinfo{year}{1996}).

\bibitem[{\citenamefont{Inouye et~al.}(2004)\citenamefont{Inouye, Goldwin,
  Olsen, Ticknor, Bohn, and Jin}}]{inouye04}
\bibinfo{author}{\bibfnamefont{S.}~\bibnamefont{Inouye}},
  \bibinfo{author}{\bibfnamefont{J.}~\bibnamefont{Goldwin}},
  \bibinfo{author}{\bibfnamefont{M.~L.} \bibnamefont{Olsen}},
  \bibinfo{author}{\bibfnamefont{C.}~\bibnamefont{Ticknor}},
  \bibinfo{author}{\bibfnamefont{J.~L.} \bibnamefont{Bohn}}, \bibnamefont{and}
  \bibinfo{author}{\bibfnamefont{D.~S.} \bibnamefont{Jin}},
  \bibinfo{journal}{Phys. Rev. Lett.} \textbf{\bibinfo{volume}{93}},
  \bibinfo{pages}{183201} (\bibinfo{year}{2004}).

\bibitem[{\citenamefont{Ferlaino et~al.}(2006)\citenamefont{Ferlaino, D'Errico,
  Roati, Zaccanti, Inguscio, Modugno, and Simoni}}]{ferlaino06}
\bibinfo{author}{\bibfnamefont{F.}~\bibnamefont{Ferlaino}},
  \bibinfo{author}{\bibfnamefont{C.}~\bibnamefont{D'Errico}},
  \bibinfo{author}{\bibfnamefont{G.}~\bibnamefont{Roati}},
  \bibinfo{author}{\bibfnamefont{M.}~\bibnamefont{Zaccanti}},
  \bibinfo{author}{\bibfnamefont{M.}~\bibnamefont{Inguscio}},
  \bibinfo{author}{\bibfnamefont{G.}~\bibnamefont{Modugno}}, \bibnamefont{and}
  \bibinfo{author}{\bibfnamefont{A.}~\bibnamefont{Simoni}},
  \bibinfo{journal}{Phys. Rev. A} \textbf{\bibinfo{volume}{73}},
  \bibinfo{pages}{040702} (\bibinfo{year}{2006}).

\bibitem[{\citenamefont{Stan et~al.}(2004)\citenamefont{Stan, Zwierlein,
  Schunck, Raupach, and Ketterle}}]{stan04}
\bibinfo{author}{\bibfnamefont{C.~A.} \bibnamefont{Stan}},
  \bibinfo{author}{\bibfnamefont{M.~W.} \bibnamefont{Zwierlein}},
  \bibinfo{author}{\bibfnamefont{C.~H.} \bibnamefont{Schunck}},
  \bibinfo{author}{\bibfnamefont{S.~M.~F.} \bibnamefont{Raupach}},
  \bibnamefont{and} \bibinfo{author}{\bibfnamefont{W.}~\bibnamefont{Ketterle}},
  \bibinfo{journal}{Phys. Rev. Lett.} \textbf{\bibinfo{volume}{93}},
  \bibinfo{pages}{143001} (\bibinfo{year}{2004}).

\bibitem[{\citenamefont{Deh et~al.}(2008)\citenamefont{Deh, Marzok, Zimmermann,
  and {Ph. W. Courteille}}}]{deh08}
\bibinfo{author}{\bibfnamefont{B.}~\bibnamefont{Deh}},
  \bibinfo{author}{\bibfnamefont{C.}~\bibnamefont{Marzok}},
  \bibinfo{author}{\bibfnamefont{C.}~\bibnamefont{Zimmermann}},
  \bibnamefont{and} \bibinfo{author}{\bibnamefont{{Ph. W. Courteille}}},
  \bibinfo{journal}{Phys. Rev. A} \textbf{\bibinfo{volume}{77}},
  \bibinfo{pages}{010701} (\bibinfo{year}{2008}).

\bibitem[{\citenamefont{Pilch et~al.}(2008)\citenamefont{Pilch, Lange,
  Prantner, Kerner, Ferlaino, N{\"a}gerl, and Grimm}}]{pilch08}
\bibinfo{author}{\bibfnamefont{K.}~\bibnamefont{Pilch}},
  \bibinfo{author}{\bibfnamefont{A.~D.} \bibnamefont{Lange}},
  \bibinfo{author}{\bibfnamefont{A.}~\bibnamefont{Prantner}},
  \bibinfo{author}{\bibfnamefont{G.}~\bibnamefont{Kerner}},
  \bibinfo{author}{\bibfnamefont{F.}~\bibnamefont{Ferlaino}},
  \bibinfo{author}{\bibfnamefont{H.-C.} \bibnamefont{N{\"a}gerl}},
  \bibnamefont{and} \bibinfo{author}{\bibfnamefont{R.}~\bibnamefont{Grimm}},
  \bibinfo{howpublished}{ar{X}iv:0812.3287v1} (\bibinfo{year}{2008}).

\bibitem[{\citenamefont{{W}ille et~al.}(2008)\citenamefont{{W}ille,
  {S}piegelhalder, {K}erner, {N}aik, {T}renkwalder, {H}endl, {S}chreck,
  {G}rimm, {T}iecke, {W}alraven et~al.}}]{wille08}
\bibinfo{author}{\bibfnamefont{E.}~\bibnamefont{{W}ille}},
  \bibinfo{author}{\bibfnamefont{F.~M.} \bibnamefont{{S}piegelhalder}},
  \bibinfo{author}{\bibfnamefont{G.}~\bibnamefont{{K}erner}},
  \bibinfo{author}{\bibfnamefont{D.}~\bibnamefont{{N}aik}},
  \bibinfo{author}{\bibfnamefont{A.}~\bibnamefont{{T}renkwalder}},
  \bibinfo{author}{\bibfnamefont{G.}~\bibnamefont{{H}endl}},
  \bibinfo{author}{\bibfnamefont{F.}~\bibnamefont{{S}chreck}},
  \bibinfo{author}{\bibfnamefont{R.}~\bibnamefont{{G}rimm}},
  \bibinfo{author}{\bibfnamefont{T.~G.} \bibnamefont{{T}iecke}},
  \bibinfo{author}{\bibfnamefont{J.~T.~M.} \bibnamefont{{W}alraven}},
  \bibnamefont{et~al.}, \bibinfo{journal}{Phys. {R}ev. {L}ett.}
  \textbf{\bibinfo{volume}{100}}, \bibinfo{pages}{053201}
  (\bibinfo{year}{2008}).

\bibitem[{\citenamefont{van Kempen et~al.}(2002)\citenamefont{van Kempen,
  Kokkelmans, Heinzen, and Verhaar}}]{vankempen02}
\bibinfo{author}{\bibfnamefont{E.~G.~M.} \bibnamefont{van Kempen}},
  \bibinfo{author}{\bibfnamefont{S.~J. J. M.~F.} \bibnamefont{Kokkelmans}},
  \bibinfo{author}{\bibfnamefont{D.~J.} \bibnamefont{Heinzen}},
  \bibnamefont{and} \bibinfo{author}{\bibfnamefont{B.~J.}
  \bibnamefont{Verhaar}}, \bibinfo{journal}{Phys. Rev. Lett.}
  \textbf{\bibinfo{volume}{88}}, \bibinfo{pages}{093201}
  (\bibinfo{year}{2002}).

\bibitem[{\citenamefont{Greene et~al.}(1982)\citenamefont{Greene, Rau, and
  Fano}}]{greene82}
\bibinfo{author}{\bibfnamefont{C.~H.} \bibnamefont{Greene}},
  \bibinfo{author}{\bibfnamefont{A.~R.~P.} \bibnamefont{Rau}},
  \bibnamefont{and} \bibinfo{author}{\bibfnamefont{U.}~\bibnamefont{Fano}},
  \bibinfo{journal}{Phys. Rev. A} \textbf{\bibinfo{volume}{26}},
  \bibinfo{pages}{2441} (\bibinfo{year}{1982}).

\bibitem[{\citenamefont{Seaton}(1983)}]{seaton83}
\bibinfo{author}{\bibfnamefont{M.~J.} \bibnamefont{Seaton}},
  \bibinfo{journal}{Rep. Prog. Phys.} \textbf{\bibinfo{volume}{46}},
  \bibinfo{pages}{167} (\bibinfo{year}{1983}).

\bibitem[{\citenamefont{Mies}(1984)}]{mies84}
\bibinfo{author}{\bibfnamefont{F.~H.} \bibnamefont{Mies}}, \bibinfo{journal}{J.
  {C}hem. {P}hys.} \textbf{\bibinfo{volume}{80}}, \bibinfo{pages}{2514}
  (\bibinfo{year}{1984}).

\bibitem[{\citenamefont{Idziaszek et~al.}(2009)\citenamefont{Idziaszek,
  Calarco, Julienne, and Simoni}}]{idziaszek09}
\bibinfo{author}{\bibfnamefont{Z.}~\bibnamefont{Idziaszek}},
  \bibinfo{author}{\bibfnamefont{T.}~\bibnamefont{Calarco}},
  \bibinfo{author}{\bibfnamefont{P.~S.} \bibnamefont{Julienne}},
  \bibnamefont{and} \bibinfo{author}{\bibfnamefont{A.}~\bibnamefont{Simoni}},
  \bibinfo{journal}{Phys. Rev. A} \textbf{\bibinfo{volume}{79}},
  \bibinfo{pages}{010702} (\bibinfo{year}{2009}).

\bibitem[{\citenamefont{Julienne and {M}ies}(1989)}]{julienne89}
\bibinfo{author}{\bibfnamefont{P.~S.} \bibnamefont{Julienne}} \bibnamefont{and}
  \bibinfo{author}{\bibfnamefont{F.~H.} \bibnamefont{{M}ies}},
  \bibinfo{journal}{J.~{O}pt.~{S}oc.~{A}m.~{B}} \textbf{\bibinfo{volume}{6}},
  \bibinfo{pages}{2257} (\bibinfo{year}{1989}).

\bibitem[{\citenamefont{{M}ies and {R}aoult}(2000)}]{mies00b}
\bibinfo{author}{\bibfnamefont{F.~H.} \bibnamefont{{M}ies}} \bibnamefont{and}
  \bibinfo{author}{\bibfnamefont{M.}~\bibnamefont{{R}aoult}},
  \bibinfo{journal}{Phys. {R}ev. {A}} \textbf{\bibinfo{volume}{62}},
  \bibinfo{pages}{012708} (\bibinfo{year}{2000}).

\bibitem[{\citenamefont{{R}aoult and {M}ies}(2004)}]{raoult04}
\bibinfo{author}{\bibfnamefont{M.}~\bibnamefont{{R}aoult}} \bibnamefont{and}
  \bibinfo{author}{\bibfnamefont{F.~H.} \bibnamefont{{M}ies}},
  \bibinfo{journal}{Phys. {R}ev. {A}} \textbf{\bibinfo{volume}{70}},
  \bibinfo{pages}{012710} (\bibinfo{year}{2004}).

\bibitem[{\citenamefont{Gao}(1996)}]{gao96}
\bibinfo{author}{\bibfnamefont{B.}~\bibnamefont{Gao}}, \bibinfo{journal}{Phys.
  {R}ev. {A}} \textbf{\bibinfo{volume}{54}}, \bibinfo{pages}{2022}
  (\bibinfo{year}{1996}).

\bibitem[{\citenamefont{Gao}(1998)}]{gao98}
\bibinfo{author}{\bibfnamefont{B.}~\bibnamefont{Gao}}, \bibinfo{journal}{Phys.
  {R}ev. {A}} \textbf{\bibinfo{volume}{58}}, \bibinfo{pages}{1728}
  (\bibinfo{year}{1998}).

\bibitem[{\citenamefont{Gao}(2004)}]{gao04b}
\bibinfo{author}{\bibfnamefont{B.}~\bibnamefont{Gao}}, \bibinfo{journal}{{J.
  Phys. B}} \textbf{\bibinfo{volume}{37}}, \bibinfo{pages}{L227}
  (\bibinfo{year}{2004}).

\bibitem[{\citenamefont{{G}ao et~al.}(2005)\citenamefont{{G}ao, {T}iesinga,
  {W}illiams, and {J}ulienne}}]{gao05}
\bibinfo{author}{\bibfnamefont{B.}~\bibnamefont{{G}ao}},
  \bibinfo{author}{\bibfnamefont{E.}~\bibnamefont{{T}iesinga}},
  \bibinfo{author}{\bibfnamefont{C.~J.} \bibnamefont{{W}illiams}},
  \bibnamefont{and} \bibinfo{author}{\bibfnamefont{P.~S.}
  \bibnamefont{{J}ulienne}}, \bibinfo{journal}{Phys. {R}ev. {A}}
  \textbf{\bibinfo{volume}{72}}, \bibinfo{pages}{042719}
  (\bibinfo{year}{2005}).

\bibitem[{\citenamefont{Fano}(1970)}]{fano70}
\bibinfo{author}{\bibfnamefont{U.}~\bibnamefont{Fano}}, \bibinfo{journal}{Phys.
  Rev. A} \textbf{\bibinfo{volume}{2}}, \bibinfo{pages}{353}
  (\bibinfo{year}{1970}).

\bibitem[{\citenamefont{Rau and Fano}(1971)}]{rau71}
\bibinfo{author}{\bibfnamefont{A.~R.~P.} \bibnamefont{Rau}} \bibnamefont{and}
  \bibinfo{author}{\bibfnamefont{U.}~\bibnamefont{Fano}},
  \bibinfo{journal}{Phys. Rev. A} \textbf{\bibinfo{volume}{4}},
  \bibinfo{pages}{1751} (\bibinfo{year}{1971}).

\bibitem[{\citenamefont{Rose}(1957)}]{rose}
\bibinfo{author}{\bibfnamefont{M.~E.} \bibnamefont{Rose}},
  \emph{\bibinfo{title}{Elementary theory of angular momentum}}
  (\bibinfo{publisher}{Wiley}, \bibinfo{address}{New York},
  \bibinfo{year}{1957}).

\bibitem[{\citenamefont{{P}ashov et~al.}(2007)\citenamefont{{P}ashov,
  {D}ocenko, {T}amanis, {F}erber, {K}n\"{o}ckel, and {T}iemann}}]{pashov07}
\bibinfo{author}{\bibfnamefont{A.}~\bibnamefont{{P}ashov}},
  \bibinfo{author}{\bibfnamefont{O.}~\bibnamefont{{D}ocenko}},
  \bibinfo{author}{\bibfnamefont{M.}~\bibnamefont{{T}amanis}},
  \bibinfo{author}{\bibfnamefont{R.}~\bibnamefont{{F}erber}},
  \bibinfo{author}{\bibfnamefont{H.}~\bibnamefont{{K}n\"{o}ckel}},
  \bibnamefont{and}
  \bibinfo{author}{\bibfnamefont{E.}~\bibnamefont{{T}iemann}},
  \bibinfo{journal}{Phys. {R}ev. {A}} \textbf{\bibinfo{volume}{76}},
  \bibinfo{pages}{022511} (\bibinfo{year}{2007}).

\bibitem[{\citenamefont{Tiecke et~al.}(2008)\citenamefont{Tiecke, Walraven, and
  Kokkelmans}}]{tiecke_private}
\bibinfo{author}{\bibfnamefont{T.~G.} \bibnamefont{Tiecke}},
  \bibinfo{author}{\bibfnamefont{J.~T.~M.} \bibnamefont{Walraven}},
  \bibnamefont{and} \bibinfo{author}{\bibfnamefont{S.~J. J. M.~F.}
  \bibnamefont{Kokkelmans}}, \bibinfo{journal}{private communication}
  (\bibinfo{year}{2008}).

\end{thebibliography}

\end{document}